\documentclass[iop]{emulateapj}

\makeatletter
\renewcommand{\bibstyle@aas}{\bibpunct{(}{)}{;}{a}{,}{,}}% 
\makeatother
\shorttitle{CERES: A New Set of Automated Routines for Echelle Spectra}
\shortauthors{Brahm et al. 2016}
\usepackage{threeparttable}
\usepackage{hyperref}
\hypersetup{colorlinks=true, urlcolor=blue, citecolor=cyan, pdfborder={0 0 0},}
\usepackage{soul}
\usepackage{natbib}

\begin{document}


\title{CERES: A Set of Automated Routines for Echelle Spectra}


\author{Rafael Brahm\altaffilmark{1,2,3}, 
Andr\'es Jord\'an\altaffilmark{1,2}, 
N\'estor Espinoza\altaffilmark{1,2}
}

% TODO: fix order and affiliation numbers.
% TODO: make sure noone is missing

\altaffiltext{1}{Instituto de Astrof\'isica, Facultad de F\'isica,
    Pontificia Universidad Cat\'olica de Chile, Av.\ Vicu\~na Mackenna
    4860, 782-0436 Macul, Santiago, Chile}
    
\altaffiltext{2}{Millennium Institute of Astrophysics, Av.\ Vicu\~na Mackenna
    4860, 782-0436 Macul, Santiago, Chile}

\altaffiltext{3}{rbrahm@astro.puc.cl, the CERES code can be downloaded from https://github.com/rabrahm/ceres}    
    
\begin{abstract}
We present the Collection of Elemental Routines for Echelle Spectra (CERES).
These routines were developed for the construction of automated pipelines for the reduction,
extraction and analysis of spectra acquired with different instruments, allowing the obtention of
homogeneous and standardised results. This modular code includes tools for handling the different steps
of the processing: CCD image reductions, identification and tracing of the echelle orders, optimal and rectangular extraction, computation of the
wavelength solution, estimation of radial velocities, and rough and fast estimation of the atmospheric parameters.
Currently, CERES has been used to develop automated pipelines for thirteen different spectrographs, namely
CORALIE, FEROS, HARPS, ESPaDOnS, FIES, PUCHEROS, FIDEOS, CAFE, DuPont/Echelle, Magellan/Mike, Keck/HIRES, Magellan/PFS
and APO/ARCES, but the routines can be easily used in order to deal with data coming from other spectrographs.
We show the high precision in radial velocity that CERES achieves for some of these instruments and we
briefly summarize some results that have already been obtained using the CERES pipelines.

\end{abstract}


\section{Introduction}

The possibility of obtaining at the same time high spectral resolution and wide spectral coverage
has made echelle spectrographs a highly demanded type of instrument. Nowadays, most
astronomical facilities count with at least one of these spectrographs \citep[see e.g.][]{vogt:1994,dekker:2000,noguchi:2002}
and they are vastly used for a wide list of astronomical applications, like the study stellar atmospheres and the search
of stellar and substellar companions by measuring radial velocity variations.
In particular, the development of echelle spectrographs has significantly raised
in the last couple of decades due to the high radial velocity precision that they can achieve with careful calibration, a
capability that has been used for discovering $\approx$ 500 extrasolar planets \citep[e.g.][]{mayor:1995,giguere:2015}.

One of the drawbacks of echelle spectrographs compared to typical spectrographs is the relative complexity demanded
in the data reduction process, due to the fact that it contains several instrumental artefacts that need to be removed in
order to extract a  wavelength calibrated spectrum amenable for astrophysical analysis.
The major complexity relies in the presence of multiple orders. These orders have quite different intensity levels due to
wavelength dependent efficiency of the instrument which can obstruct their identification in some cases. Additionally,
these orders have in general a significant curvature which has to be taken into account during the extraction and in
some particular cases contiguous orders tend to overlap each other in the vertical direction which difficults a proper estimation
of the scattered light. Moreover, echelle spectrographs can contain additional calibration fibres and image slicers that
further complicate the processing of the data.
An important fraction of current echelle spectrographs have their own reduction pipelines specifically designed for the properties of
each instrument \citep[e.g.][]{bochanski:2009,mink:2011}, while in other particular cases there is no dedicated pipeline at all. This fact can produce
some inconsistencies when spectra obtained from different instruments are used in the same analysis, in particular
when the reduction steps include human intervention. Even in the case of working with data of a particular telescope,
the automatisation of the data processing is desirable and is specially crucial when working on obtaining precision radial velocities at different epochs, because slight
changes in the reduction steps can introduce significant systematic effects that propagate to the estimation of the Doppler shifts. 

There have been already some attempts to develop computational tools for the automated processing of data originated from 
different echelle spectrographs, but their use has not been extended for more than a couple
of instruments so far. For example, \citet{mills:2003} presented the open source code ECHOMOP, which has been mostly used in the
processing of data obtained from the Utrecht Echelle Spectrograph of the 4.2-m William Hershell Telescope, while \citet{sosnowaska:2015}
presented the flexible reduction library for the ESPRESSO project which is also able to process data from HARPS and HARPS-N.
Along the same line, the MIDAS system \citep{banse:1983} developed by ESO included a package designed to process echelle spectra \citep{ballester:1992},
which has been used to develop pipelines for most of ESO spectrographs.

In this paper we present a new set of computational routines for developing fully automated reduction pipelines
for data of echelle spectrographs. This modular code called CERES (\textbf{C}ollection of \textbf{E}lemental \textbf{R}outines 
for \textbf{E}chelle \textbf{S}pectra) is mostly written in Python, but contains also  C and Fortran routines when speedy execution 
demands it. We have developed reduction pipelines for thirteen different spectrographs, and these recipes can be used as a guide for 
building pipelines for other instruments. The principal aim of the pipelines that we have developed is the handling of low signal to 
noise ratio data and the measurement of precision radial velocities in the context of extrasolar planets.

In \S~\ref{structure} we describe the structure of the CERES echelle pipelines and the corresponding variations for each type of spectrograph.
In \S~\ref{instruments} we list the instruments that are currently supported by CERES, while in \S~\ref{results} we discuss the performance of
some of the CERES pipelines. Finally, in \S~\ref{concl} we summarise our work.

\section{Structure and reduction steps}
\label{structure}
\subsection{General considerations}

The main purpose of the CERES routines is the development of completely automated pipelines that are able to generate
optimally extracted, wavelength calibrated and instrumentally corrected spectra, plus additional parameters like radial velocities,
bisector spans and stellar atmospheric parameters. The input is raw images of echelle spectrographs, and we aim to fully avoid the need for human intervention in the process.
CERES routines have been successfully implemented for handling data of echelle spectrographs with quite different
specifications, including fibre-fed and slit spectrographs. A description of the particular functions used by CERES can be found
in the github repository\footnote{}
%Due to their higher stability, the pipelines designed for fibre-fed spectrographs are usually easier to implement
%and more robust than the ones we have implemented for slit spectrographs. However, we have been successful in developing
%CERES pipelines for both types of instruments.

All CERES pipelines have the same general structure. They have a main code that drives all the steps required to obtain a
reduced and analysed spectrum. The main code can call functions from a general module that contains tasks that can be
used by different pipelines (\texttt{GLOBALutils}), but it can also call functions from another \texttt{Python} module
that contains tasks specifically designed for a particular instrument. In this way, the different CERES pipelines mostly use
the same functions in which the parameters change according to the instrument specifications. However, there are some
specific functions for which the operations can be structurally different.
Most of the tools are coded in \texttt{Python} but there are also some time-consuming tasks for which we wrote code in
\texttt{C} but which are wrapped to be called directly from \texttt{Python}.
Some of the time consuming tasks have been also paralelized to further accelerate the reduction process.
Even though all pipelines share a similar structure, there are several particular differences due to the differing properties of each instrument.

In order to simplify the description of the CERES routines we will briefly explain the structure of a typical echelle image and we will define some useful concepts.
The high resolving power obtained with echelle spectrographs is achieved thanks to the high incidence angle used between the incident light and the plane grating.
In oder words, echelle spectrographs work at very high spectral orders. However, this configuration of the instrument produces that, after the beam is dispersed by the grating,
different orders of the spectrum overlap with each other. In order to correct for this effect, a second dispersion element is used which disperses the light in the perpendicular direction
with respect to the grating and in this way separates the spectral orders. This operation allows also to fit the complete spectrum in a single rectangular CCD, which will register a set of 
traces corresponding to the different echelle orders. For the rest of the paper we will refer to the direction in which the beam is dispersed by the plane grating as the \textit{dispersion direction},
while the direction in which the orders are separated with the second dispersion element will be called \textit{cross-dispersion direction}. Moreover, for simplicity we will arbitrarily consider that the
dispersion direction (cross-dispersion direction) goes in the horizontal (vertical) direction of the CCD or along its rows (columns).

\begin{figure*}
\plotone{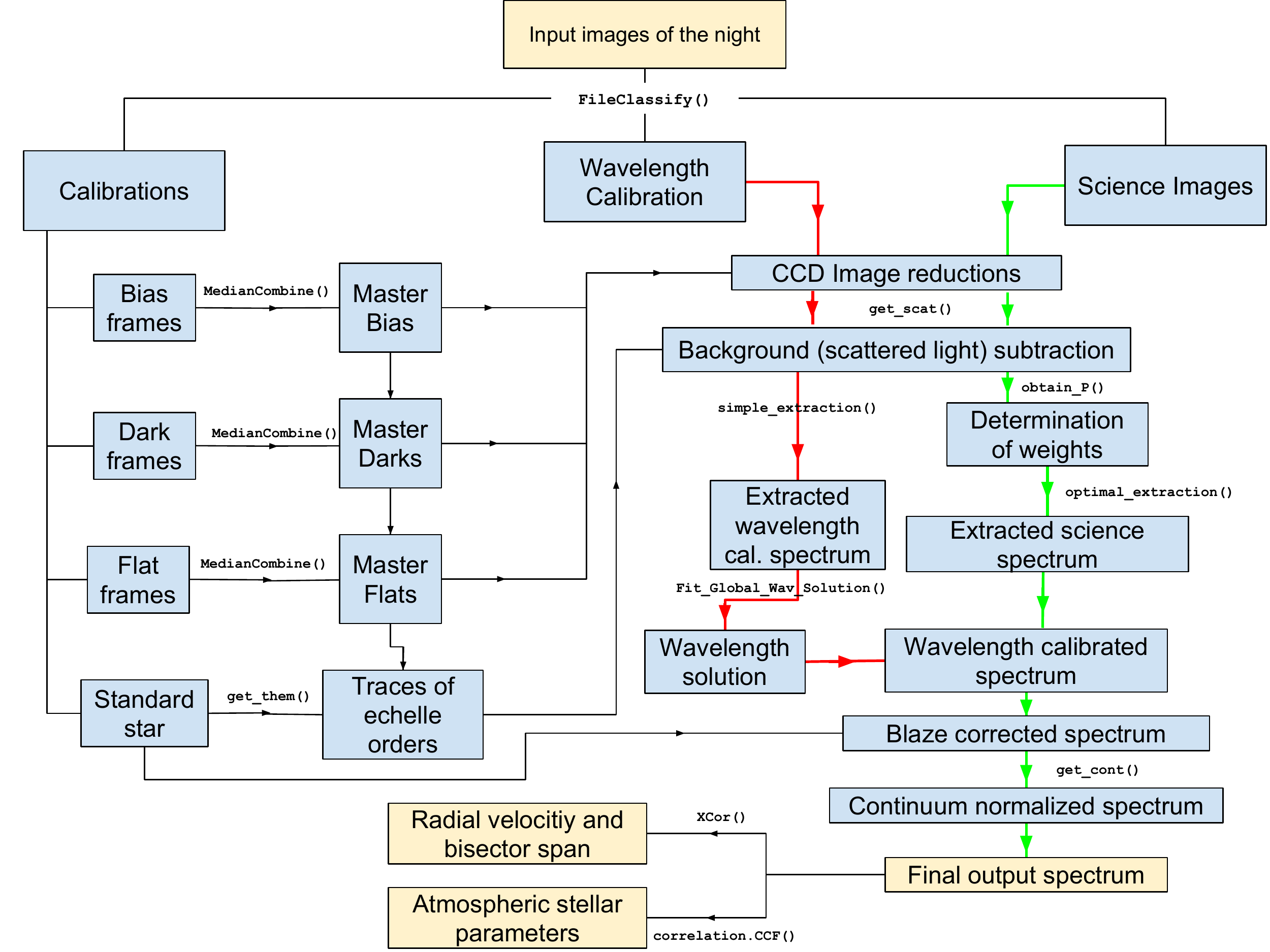}
\caption{Flow diagram showing the images and processes that are in general used by the CERES pipelines for slit spectrographs. The most important functions at each step are also shown.
 \label{slit}}
\end{figure*}

\begin{figure*}
\plotone{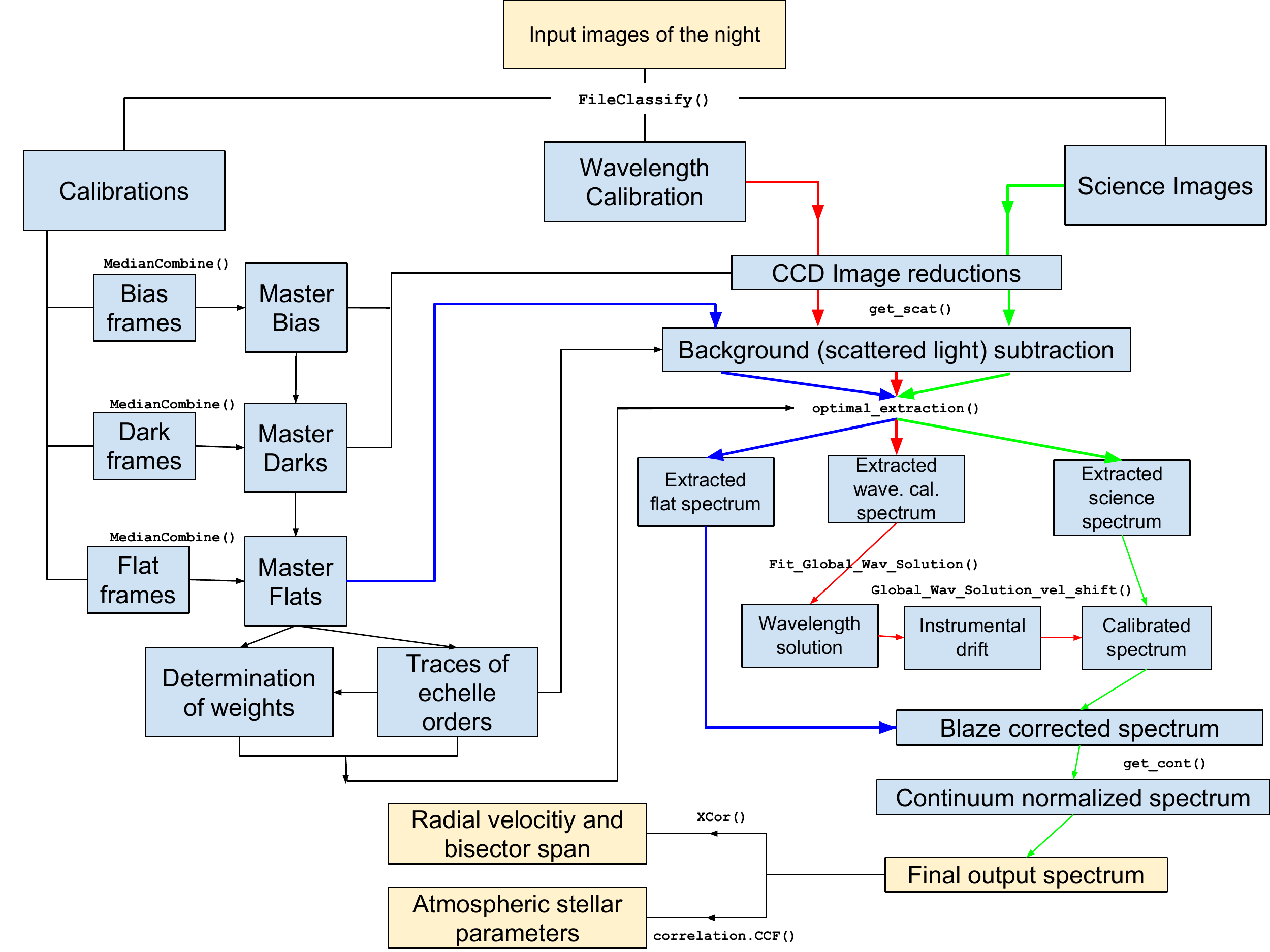}
\caption{Flow diagram showing the images and processes that are in general used by the CERES pipelines for fibre-fed spectrographs. The most important functions at each step are also shown.
 \label{fibre}}
\end{figure*}

\subsection{Pre-processing}
\subsubsection{Classification of frames}

The first step of any CERES pipeline is to identify all the images that are going to be used in the reduction process
and to classify them according to their types. CERES receives as input the path to the directory containing all the
raw images. Given that there is no unified header keywords, a particular function has to be defined for each
spectrograph that performs the classification. Common image types are: bias frames, dark frames, flat frames,
wavelength calibration frames and science images. CERES reads the header of each image and according
to the corresponding keywords of each instrument, proceeds to classify the images by creating lists for the
different data types that contain the full names of the corresponding raw images.

\subsubsection{Master CCD Frames}
\label{master}
Once the classification of the images is done, the next step is to construct the master CCD calibration images
by median combination. The exact type of required calibration images strongly depends on the specifications
of each instrument. Bias and dark master frames are usually not included as standard calibrations for several
spectrographs but these master frames are generated if required. In the case of master dark frames, if there is no
dark that matches the exposure time of a particular science image, the CERES routines interpolate the
intensity of the master dark frames in order to generate one with the required exposure time. The correction of
the science images is performed by subtracting the master bias frame and the corresponding master dark frame.

Flat-field correction is a relatively complex procedure in the case of echelle spectrographs. This procedure is required
to correct for sensitivity variations of the detector pixels. The use of a classical flat-field in which
the full detector is illuminated by an homogeneous source is not reliable in spectroscopy because the sensitivity of each
pixel depends also on the particular wavelength that it receives. For this reason, the light of the source used for the flat-field
correction must be also dispersed by the instrument. In the case of fibre-fed echelle spectrographs, it is physically imposible to
illuminate the full detector. However, in the case of slit spectrographs, a common procedure is to obtain
the spectrum of a continuum source but using a longer slit than the one used for obtaining the science spectra.
Spectrographs like Magellan/MIKE \citep{bernstein:2003} and Keck/HIRES \citep{vogt:1994} follow this calibration
procedure which ensures that the sensitivity variations of the CCD in the borderline regions of the echelle orders of the
science spectra are properly corrected. We will name these type of calibration image  ``long-slit flat frames" from now on.
What CERES actually does with long-slit flats is first to co-add them in order to generate a master long-slit flat frame. Then,
after having identified and traced the echelle orders (see \S~\ref{sec:orders}), CERES
selects the slices of this master flat that contain the echelle orders, and a 2D median filter is computed for each slice.
Finally, each of the slices containing the echelle orders is divided by its corresponding median filtered surface and
at the same time the inter-order regions of the image are filled with values equal to one. This normalised long-slit
flat frame will contain the high frequency pixel sensitivity of the CCD with its corresponding wavelength dependence.
After subtracting the bias and dark frames, the science images will be corrected by dividing them by the normalised
long-slit flat frame. This procedure only corrects for the high frequency pixel variations. Some other slit spectrographs
use a different approach for performing the flat-field correction. For example, for the echelle at the DuPont telescope
a diffuser is placed in front of the spectrograph which spreads the dispersed light of the afternoon sky and generates
a smooth and homogeneously illuminated image which conserves also the approximate wavelength dependence at
each pixel. This type of image is known as miky-flat. In this case, CERES co-adds the miky-flat images in order
to generate a master milky flat and then a 2D median-filtered frame is obtained of the whole image. Finally, the
master milky-flat is divided by the median-filtered frame to generate the normalised milky-flat, for which the inter
order regions are not modified because in this case they contain real signal. The science frames are then corrected
by dividing them by the normalised milky-flat after the master bias and dark subtraction.

In the case of fibre-fed spectrographs, the flat-field correction is not attempted, which is due mostly due to the
complexity of illuminating with high intensity the border regions of each echelle order. Pixel sensitivity variations for fibre-fed 
spectrographs are only partially corrected following another approach which is explained in \S~\ref{final}. This procedure
requires images in which the spectrum of a continuum lamp in registered. This images will be called``fibre-flat frames" hereon. CERES co-adds all the fibre-flats to generate the master fibre-flat, which is then used to identify
and trace the orders (see \S~\ref{sec:orders}), compute the weights for the optimal extraction (see \S~\ref{final}), and
correct for the blaze function (see \S~\ref{sec:orders}).

\subsubsection{Identification of the echelle orders}
\label{sec:orders}
\begin{figure*}
\plotone{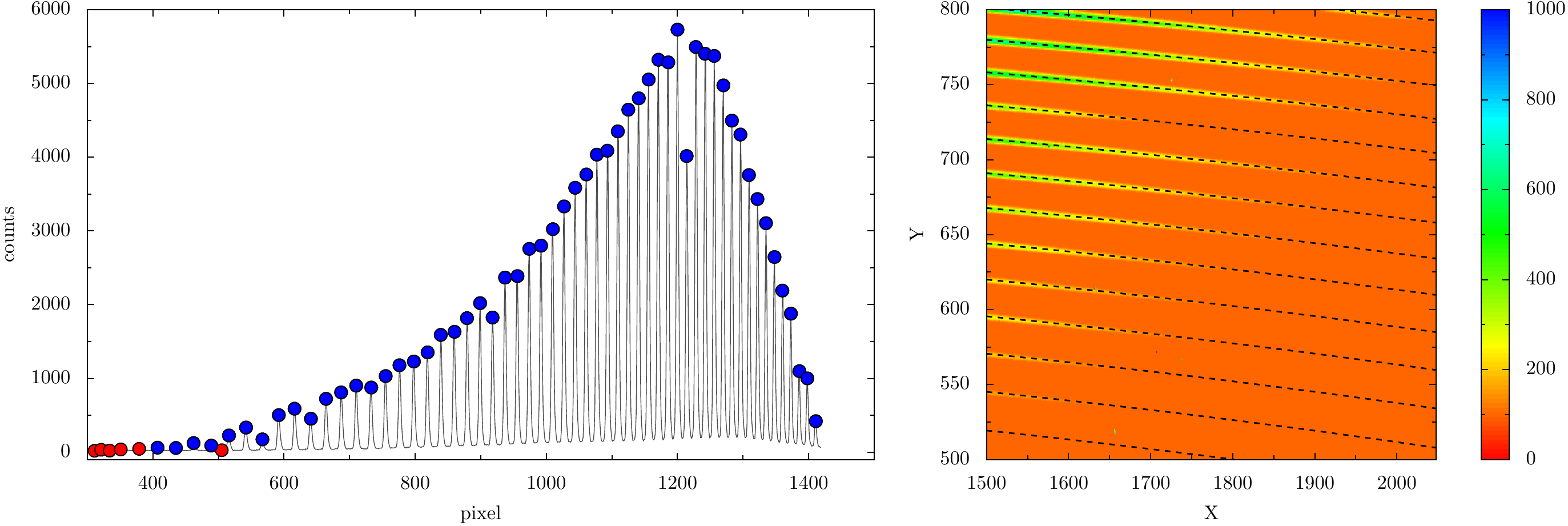}
\caption{Left: vertical cut of an image of the DuPont/Echelle. The blue circles correspond to the orders identified by our algorithm while the red circles are the other maxima that are rejected. Right: portion of an echelle image used to determine the traces of the echelle orders which are displayed as dashed lines. Given that the algorithm uses the information of the complete image to trace each order, good performance is achieved even in region with low signal-to-noise ratio.
 \label{traces}}
\end{figure*}

One key step in the processing of echelle data is the identification and tracing of the echelle orders.
In the case of fibre-fed spectrographs, an image of the spectra of a continuum lamp is used to
perform this procedure, while a spectrum of a bright object is usually used for slit spectrographs.
In order to find the orders, the central columns of the image are used because usually they have
 higher SNR than the other zones. The exact number of columns is a free parameter that can be modified by the user
but usually is of the order of 10 pixels. These columns are median combined along the  dispersion
direction for constructing a reference vertical cut of the CCD without cosmic rays or cosmetic artefacts.
Then, this reference column is convolved with a Gaussian kernel for smoothing it, where the width of
the Gaussian is another parameter that can be modified by the user. All the peaks of the
smoothed reference column are identified and then an iterative procedure is applied to reject shallow
peaks that have smaller counts than $N$ times the dispersion of the counts in the inter-order zones,
where $N$ is another adjustable parameter. The left panel of Figure~\ref{traces} shows the vertical cut in an image
obtained with the echelle spectrograph mounted on the DuPont telescope that is used to identify the
orders.

As opposed to common spectrographs, the traces of most echelle spectrographs have a strong curvature on the CCD.
Starting from the central position of each order computed in the last step, the vertical position of the order in
each column is identified by fitting a Gaussian in zones centered on the positions contiguous to the already identified order centers. 
The width of the zone where the Gaussian is fit must be similar to the approximate cross-dispersion (vertical) extension of the orders. In order to avoid errors in the tracing of the orders arising from low signal to noise, cosmic rays, or instrumental artifacts, a procedure is applied in which the drift of the orders between the new column
and the previous column is computed and positions that show drifts 3 times greater than the
dispersion of the already computed drifts for that order are rejected and the centroid for those positions are replaced by the one of the contiguous position plus the median drift.

Finally a high order polynomial is fitted to the centroids computed for each order and the coefficients of the fit
are saved for extracting the science and calibration spectra. The right panel of Figure~\ref{traces} shows a portion
of a DuPont image and the traces identified by the CERES algorithm.

Some slit spectrographs mounted directly on the telescope can suffer from strong flexures, which
produce as a result that the vertical position of the orders can significantly change depending
on the pointing position of the telescope.
To handle this effect, CERES includes a function to retrace the orders by using the reference traces
and performing a cross-correlation against the science images in pixel space. The pixel
displacement of the maximum of the cross-correlation function is taken as the
instrumental drift and all the reference traces are displaced by that amount.

\subsubsection{Scattered light subtraction}

In echelle spectrographs, in addition to the dispersed light registered in each order, some scattered light generated
by the echelle grating and the roughness of the optical surfaces is also detected by the CCD.
This contamination produces a smooth background that has to be removed before extracting the spectra
in order to conserve the true depth of the absorption lines. CERES uses a simple algorithm to correct for
this effect (\texttt{get\_scat}). At this stage, the traces of the orders and their widths are already known, and this can  be used to
select only the inter-order zones of the image. For each column, the algorithm computes the median flux in each
inter order region and then a linear interpolation of these values in the vertical (cross-dispersion) direction is performed
to infer the scattered light level in the intra-order regions of the image. After performing this process for each column we will
end up with a two dimensional map of the scattered light. In order to smooth the effects of spurious signals, a two dimensional
median filter is applied to the constructed scattered light surface, and finally the filtered image is subtracted from each original
science image.

\subsection{Extraction}
\label{sec:ext:ceres}

The extraction of a spectrum refers to the process of adding up all the signal around the trace in the direction perpendicular
to the dispersion, going thus from a 2D image to a 1D spectrum. In the case of an echelle spectrograph the extraction process
produces a 1D spectrum for each order. Before adding up the signal, all the systematic effects must be removed, which means
that the master calibration frames must be applied to the science image, bad columns must be corrected and the
scattered light background must be subtracted as described above.

CERES contains two algorithms for performing the extraction. The simpler one corresponds to the simple
sum of the flux of all the pixels contained in a vertical window of width defined by the user. This algorithm is known
as ``rectangular extraction". The second algorithm that CERES implements is called optimal extraction \citep{horne:1986}, which is
particularly useful to obtain high quality spectra from low SNR data. This algorithm relies on the determination of the appropriate weights 
across the profile of the objects that produce the minimum expected variance for the expected value of the total flux, while keeping the 
estimator of the flux unbiased. If we denote these weights by $W_{x,\lambda}$, where $x$ is the $x$th pixel in the direction perpendicular 
to the dispersion direction and $\lambda$ is the $\lambda$th pixel in the dispersion direction, then the optimal extracted flux can be 
written as
\begin{eqnarray*}
F_\lambda = \sum_{x_1}^{x_2} W_{x,\lambda} F_{x,\lambda},
\end{eqnarray*}
where $F_{x,\lambda}$ is the measured flux in the CCD in the $x$th and $\lambda$th pixel, and $x_1$ and $x_2$ are the limits of the 
vertical window whose width is defined by the user (note that, for rectangular extraction, $W_{x,\lambda} = 1$). The idea of
\cite{horne:1986} is to express these weights in terms of the fraction of the flux of the object in the $x$ direction, 
$P_{x,\lambda} = F_{x,\lambda}/\sum_{x' = x_1}^{x'=x_2}F_{x',\lambda}$ (note that this constrain implies that $\sum_{x = x_1}^{x=x_2} P_{x,\lambda} = 1$). 
By imposing the minimum variance and unbiasedness conditions on the retrieved flux, it is easy to show that
\begin{eqnarray*}
W_{x,\lambda} = \frac{P_{x,\lambda}/\textnormal{Var}(F_{x,\lambda})}{\sum_{x_1}^{x_2} P_{x,\lambda}^2/\textnormal{Var}(F_{x,\lambda})},
\end{eqnarray*}
where $\textnormal{Var}(F_{x,\lambda})$ is the variance of the flux in the $x$th and $\lambda$th pixel which in this model 
is given by $F_\lambda P_{x,\lambda}/G + \textnormal{RON}/G^2$, where $G$ is the detector gain and $\textnormal{RON}$ is the readout noise. 
In order to obtain the weights, the original algorithm of \cite{horne:1986} assumes the $P_{x,\lambda}$ are smooth functions in $\lambda$ and, 
thus, modelled them as low order polynomials in $\lambda$ subject to being normalized in the $x$ direction (i.e., $\sum_{x=x_1}^{x=x_2} P_{x,\lambda} = 1$). 
However, echelle spectra are typically highly distorted in the $x$ direction as well. We thus follow \cite{marsh:1989} who extends this idea 
to deal with highly distorted spectra by modelling the $P_{x,\lambda}$ not in the $x$ direction, but in the actual \textit{traces} of the 
spectra. The strategy deals with the problem by fitting $K$ different polynomials positioned parallel to each other along the traces and 
separated by a distance $S$. 

The number of polynomials, $K$, must be higher than the width in pixels of the order and can be modified by the user. The distance $S$ is 
automatically calculated using the width in pixels and this number of polynomials. The polynomials are computed in an iterative process 
which allows to identify cosmic rays and correct for them. In the case of stable fibre-fed spectrographs, the weights are computed from 
a calibration image known as a fibre flat, which corresponds to a spectrum of a continuum lamp. In the case of slit spectrographs, the 
weights are computed from the same science images that are being extracted.
%}

\subsection{Wavelength Calibration}
\label{sec:wav:ceres}
\subsubsection{ThAr lamps}

The most common procedure for calibrating the science spectra in wavelength is to use the spectrum
of a reference lamp filled with a particular gas. When this gas is heated it radiates only in certain
narrow emission lines according to the particular allowed electronic transitions of the atoms present in the gas.
If the lamp has been characterized, it can be used to generate a mapping between the pixel position and
the corresponding wavelength. This type of light source is commonly known as arc lamp, and depending
on the particular wavelength range of interest, the arc lamp can contain different combination of gases
(H, He, Ne, Ar, Na, Cu, Hg). In the case of high resolution optical echelle spectrographs the most
commonly used calibration lamp nowadays is one composed of Thorium (Th) and Argon (Ar).

The extraction procedure adopted for the spectra of the ThAr lamps depends on the type of spectrograph used.
For fibre-feb spectrographs the extraction is performed exactly as it is performed for stellar spectra, which means
that the weights determined from the flat frames are used by the optimal extraction algorithm. In the case of slit
spectrographs, due to the absence of a reference profile for determining the weights, the simpler rectangular
extraction method is used.

Once the spectra of the ThAr lamp have been extracted, the process to compute the wavelength solution
is the same one for every spectrograph. Nontheless, for each spectrograph a specific reference line list is required.
The exact format of this set of reference lines is  as follows. For each echelle order, the pipeline requires
the existence of a text file in the $wavcals$ directory which is located inside the directory containing the
particular pipeline. These files must contain at least two columns, the first one containing the approximate pixel position
of the ThAr lines present in the corresponding echelle order (${x_i^{ap}}$), while the second column should contain the associated
wavelength value of the emission line (${\lambda_i}$).

Given that for some spectrographs the pixel position of the emission lines can drift significantly on a timescale on
the order of days, months or years, the pipeline first computes a rough estimation of this long-term instrumental
drift in the dispersion direction measured in pixels ($\Delta_p$).  $\Delta_p$ is obtained by computing the
order-by-order cross-correlation function between the corresponding extracted ThAr spectrum and a binary
mask. This binary mask is constructed from the reference text files, where it takes values equal to one in the regions
containing emission lines, and values equal to zero elsewhere. The cross-correlation functions for all the orders
are combined, and the position of the maximum of the summed cross-correlation function is assumed as $\Delta_p$,
which is then applied to shift the approximate pixel positions of all the emission lines defined in the reference text files.
For stabilised spectrographs like HARPS, FEROS, or Coralie, this procedure is not strictly required because
their long-term instrumental drifts are smaller than one pixel.

The next step performed by the pipeline is to determine a more precise value of the pixel position of each emission line
for the analysed ThAr spectrum. To achieve this goal, Gaussian functions are fitted to the extracted ThAr spectrum
in zones around ${x_i^{ap}}$. The mean of each Gaussian is used as the precise pixel position of the emission lines $x_i$.
After this procedure each emission line $i$ possesses a wavelength value ($\lambda_i$), a precise pixel position ($x_i$)
and an echelle order ($j_i$), which extends from 0 to the number of detected orders minus 1, ordered from the reddest to
the bluest.
In order to achieve the highest precision possible CERES routines have a careful treatment of zones with
blended emission lines. These zones can be defined in the reference text file, and the algorithm will fit multiple
Gaussians in those regions of the spectrum.

The next step consists in fitting iteratively a polynomial between $\{x_i\}$ and $\{\lambda_i\}$
for each echelle order $j$. In tis process CERES rejects strong outliers that correspond to poorly identified emission lines.
This procedure delivers also the approximate wavelength value of the central pixel of each echelle order $j$ ($\lambda^{c}_j$),
which can be used to determine the $real$ order numbers $\{m_j\}$ of the instrument\footnote{By ``real" order numbers we refer
to the physical order that corresponds to the one appearing on the grating equation.}. In practise, we search for the integer $m_0$
such that
\begin{equation}
m_0 + j = m_j.
\end{equation}
The grating equation states that $\lambda \propto m^{-1}$ and therefore the correct $m_0$ value will be the one
that produces the smaller slope of the following equation:
\begin{equation}
y(j) = (m_0 + j) \lambda^c_j.
\end{equation}
Once $m_0$ has been determined, each emission line will have also its corresponding $real$ echelle order $m_i=j_i+m_0$.

Once the {\em real} order numbers are known, the pipeline computes a global wavelength solution
in the form of an expansion of the grating equation \citep[see Section 2.6 in][]{baranne:1996} using Chebyshev polynomials.
This fit includes also a $3\sigma$ iterative procedure in which more outlier emission lines are rejected.

\begin{figure}
\plotone{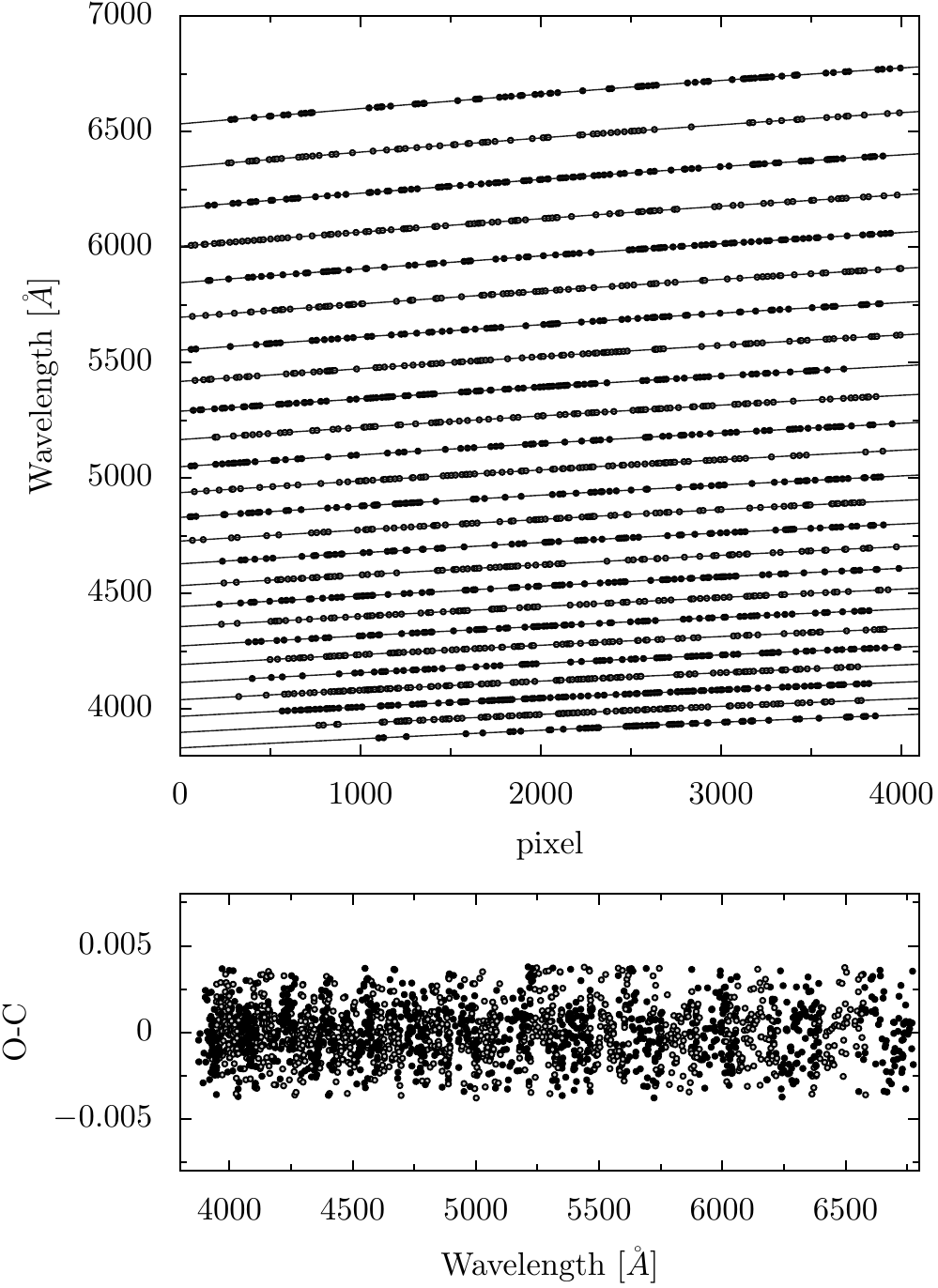}
\caption{Top panel: global wavelength solution that produces the best fit to the positions of the emission lines of a ThAr lamp observed with the FEROS spectrograph. The circles correspond to the wavelength of the particular emission lines that are used in the fit. The solid line is the fitted global solution.
Bottom panel: residuals produced with the global wavelength solution of the top panel.
 \label{wavesol}}
\end{figure}

%until the root-mean-square of the global
%solution\footnote{The solution is global in the sense that it applied to the whole CCD as opposed to a single order.} is below a
%certain threshold which mostly depends on the instrumental resolving power
%\footnote{Formally, the threshold we use is $\frac{3 \times 5000}{100 R} \AA$, where $R=\Delta\lambda/\lambda$ is the resolving
%power.}.
In detail, our global wavelength solution takes the form
\begin{equation}
\label{wavsol}
\lambda (x,m) = \frac{1}{m} \sum_{i=0}^{n_m} \sum_{j=0}^{n_x} a_{ij} c^i (m) c^j (x),
\end{equation}
where $x$ and $m$ refer to the pixel value and echelle order number, respectively, $c^n$ denotes the
Chebyshev polynomial of order $n$, $a_{ij}$ are the coefficients that are fitted to obtain the wavelength
solution, $n_m$ is the degree of the Chebyshev  polynomial in $m$ and $n_x$ is
the degree of the Chebyshev  polynomial in $x$. The values of $n_m$ and $n_x$ will depend on the
particular properties of each instrument and for the our set of pipelines  they were determined by visually
inspecting if there was any structure in the residuals of the wavelength solutions. 
Figure~\ref{wavesol} shows an example of a global wavelength solution of the FEROS spectrograph.
The bottom panel of that figure shows that the residuals in the wavelength
position of the ThAr emission lines are in general below 0.003 \AA.

The reference wavelength solution of a given night is usually obtained with the calibration frames taken in the
afternoon and posterior ThAr spectra are obtained during the night to monitor any instrumental drifts of the system.
In the case of unstabilised instruments like the ones mounted on the telescope or the ones that are not
controlled in temperature and pressure, additional ThAr exposures must be obtained before and after
each science image in order to obtain the best RV precision achievable. For stabilised instruments like
HARPS, CORALIE and FEROS the acquisition of ThAr spectra during the night may be not mandatory.
However, in the case of these instruments, a second fibre is always available for obtaining a simultaneous
ThAr spectrum while observing the science target to trace subtle instrumental drifts.

In order to determine the velocity drift of a given ThAr spectrum with respect to the ThAr spectrum that
was used to compute the reference wavelength solution, the pipeline first identifies the set of 
$\{x_i\}$, $\{\lambda_i\}$ and $\{m_i\}$ values for the new ThAr spectrum following the same steps
described in the previous paragraphs. However, then, instead of searching for new Chebyshev
coefficients by fitting equation \ref{wavsol}, the pipeline assumes that the wavelength solution
should have been modified just by a multiplicative Doppler factor, and therefore the $a_{ij}$ coefficients
of the reference wavelength solution are held fixed and the set of $\{x_i\}$, $\{\lambda_i\}$ and $\{m_i\}$
values are fitted to the following model:
 \begin{equation}
\lambda (x,m) = \frac{1}{m} \left( 1 + \frac{\delta v_{inst}}{c} \right) \sum_{i=0}^{n_m} \sum_{j=0}^{n_x} a_{ij} c^i (m) c^j (x),
\label{deltvinst}
\end{equation}
which contains only a single free parameter $\delta v_{inst}$ that corresponds to the velocity Doppler
shift between both ThAr spectra.

\subsubsection{Fabry-Perot system}
Even though the use of ThAr lamps allows to achieve very precise wavelength solutions that enable the computation
of precise RVs, they are quite far from being perfect calibrators. Only few emission lines are generally visible in the
bluer ($\lambda<4000\ \AA$) and redder ($\lambda>7000\ \AA$) parts of the wavelength coverage. The intensities of
the lines are not uniform, which produces the saturation of some lines in some orders in order to allow the detection of fainter ones.
In some cases when the simultaneous calibration lamp is used, the saturation of some ThAr lines can even contaminate
contiguous orders dedicated to obtain the science spectra. Moreover, the internal composition of the lamps
can change, which can produce long term instrumental drifts in the measured RVs. Also, they do expire at some point
and some subtle manufacturing differences in the lamps can induce other systematic errors when a new lamp is installed.
A couple of alternative wavelength calibration systems have been developed with the goal of replacing the
use of the ThAr lamps. The most promising one is the laser frequency comb \citep{li:2008}, which produces a forest
of emission lines with similar intensities whose exact wavelength positions can be synchronized with
radio-frequency oscillators referenced to atomic clocks. However, the cost of such systems is rather high and they are still under study and development. An alternative approach is the use of Fabry-Perot interferometers \citep{halverson:2014, reiners:2014, sarmiento:2014, sturmer:2016}, in which a pair of parallel reflecting
surfaces that are illuminated with a continuum source produce a fringed pattern due to the constructive and
destructive interference that arises from the interaction of the light reflected in both surfaces. The spectra produced
by a Fabry-Perot system are populated with a forest of emission lines that extend across the full wavelength coverage of
the spectrograph and where the separation between two contiguous lines depends on the effective distance
of the two reflective plates. With this type of systems, the long term precision of the wavelength calibration
does not depend on the source lamp and only on the properties of the reflective plates.
%Figure \ref{fig:fp}
%compares two spectra in an echelle order, one obtained with a ThAr lamp and the other
%obtained with a Fabry-Perot system.

One of the major drawbacks of Fabry-Perot interferometers as wavelength calibration systems is that they do not
deliver directly an absolute wavelength calibration as opposed to ThAr lamps where the characteristic pattern of
the spectrum can be used to associate specific wavelengths to each line. What is commonly done in spectrographs
like CORALIE is to compute the absolute wavelength calibration with the spectrum of a ThAr lamp obtained in the
afternoon calibrations. Inmediately after that, the absolute solution is then improved with the acquirement of Fabry-Perot spectra.
The instrumental drifts during the night are calculated then by the acquirement of Fabry-Perot spectra with the comparison fibre.
With this procedure the ThAr lamp is only marginally used, extending its useful life, the wavelength solution is determined
with higher accuracy and the science observations do not suffer from the contamination of saturated ThAr lines.

CERES includes the functions required for handling Fabry-Perot caibration spectra and they have been added to the pipeline that handles
data from the CORALIE spectrograph. In this case, for computing the wavelength calibration of a science spectrum,
3 images are required: the ThAr lamp at both fibres (TH2), the ThAr lamp at the object fibre
while the Fabry-Perot is at the comparison fibre (THFP); and the stellar spectra at the object fibre while the Fabry-Perot
is at the comparison fibre (OBFP). The global absolute wavelength solution is computed for both fibres with the reduced
spectra of TH2. Then, the Fabry-Perot spectra of the comparison fibre in THFP is calibrated in wavelength by the
computation of $\delta v_{inst}^1$ between the ThAr spectra of the TH2 and THFP images using equation \ref{deltvinst}.
Finally the instrumental drift $\delta v_{inst}^2$ between the THFP and the OBFP images is calculated via the
computation of the cross-correlation function between both Fabry-Perot spectra.

\subsection{Final output and post-processing}
\label{final}
\subsection{Final output}

\begin{figure}
\plotone{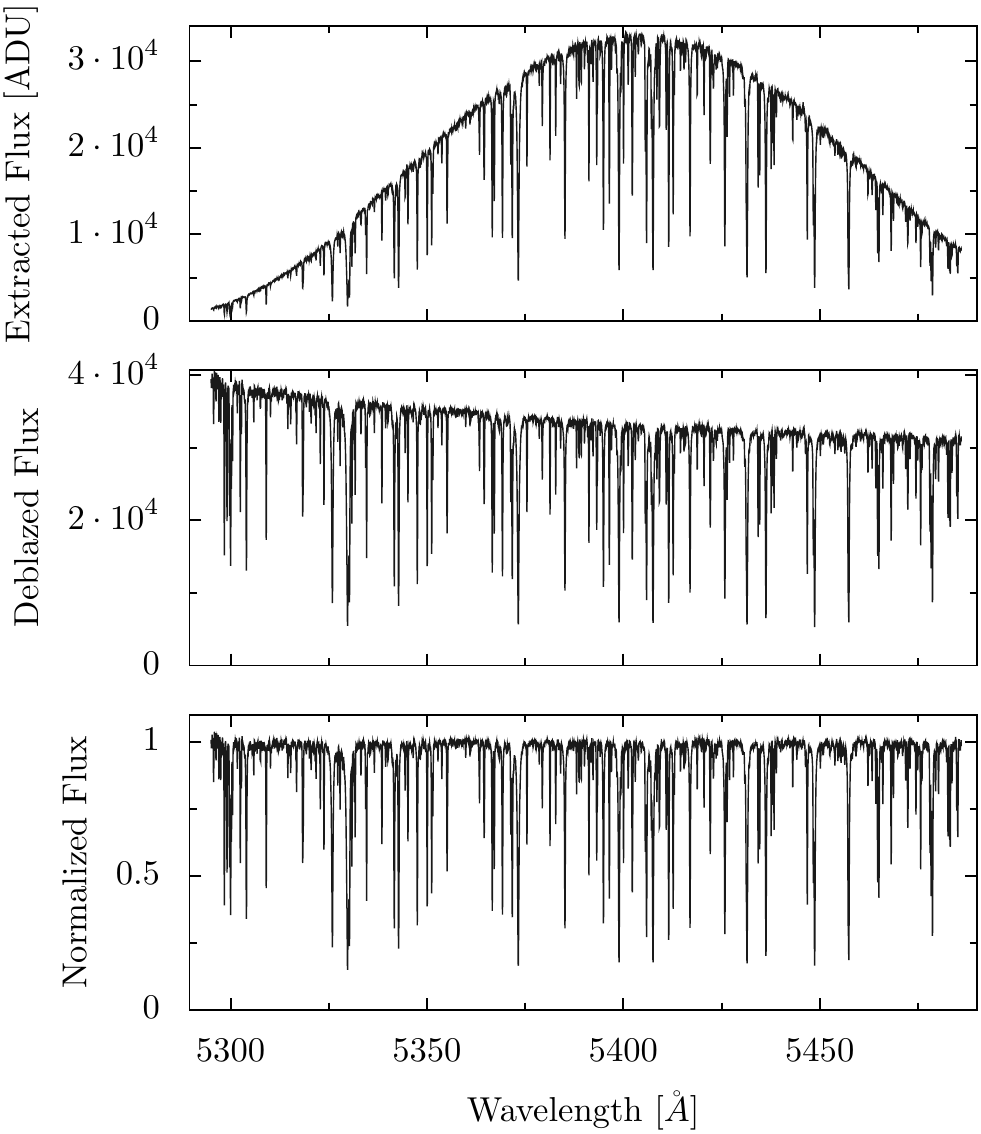}
\caption{Top panel: Optimally extracted FEROS spectrum of the star HD72673 for a particular echelle order. Central panel:
Extracted spectrum divided by the blaze function, which is obtained for the flat frames in this case. Bottom panel: continuum normalised spectrum obtained with by a polynomial fitting of the debased spectrum.
 \label{output}}
\end{figure}

Several outputs are produced after each pipeline is executed. The reduced spectrum is saved in a three
dimensional \texttt{fits} file with the following general form: \texttt{[data type,echelle order,pixel]}. The \texttt{data
type} dimension has generally eight entries. The first entry (\texttt{[0,:,:]}) is the matrix containing the wavelengths
of each pixel for each echelle order after computing all the instrumental velocity drifts.
The second entry (\texttt{[1,:,:]}) corresponds to the optimally extracted stellar flux.
The top panel of Figure~\ref{output} shows the extracted flux of a G-type star for one
order obtained with the FEROS spectrograph.
The third entry contains a measure of the error associated to the extracted flux (Inverse variance).

The fourth entry corresponds to the deblazed stellar flux (central panel of Figure~\ref{output}) and
the fifth entry contains its associated inverse variance.
In echelle spectrographs, the flux of each order registered by the CCD contains a strong instrumental
modulation which has to be corrected in order to be analysed. This instrumental effect is known as the
blaze function and it is produced due to interferences in the echelle grating, and therefore it depends on
its blaze angle.
There are several approaches that can be followed to correct for the blaze function. The complexity of
the chosen method depends on the purpose that the science spectra will have. 
Given that the main purpose of the CERES pipelines is the use of the narrow stellar absorption
lines for computing precision radial velocities and the determination of stellar atmospheric parameters, 
the large scale modulations of the spectrum do not present a fundamental problem.
For this reason, CERES uses simple procedures for estimating the blaze function.

In the case of fibre-fed spectrographs, the blaze function of the science spectra is partially corrected
by using the \textit{fibre flat frames} described in \S~\ref{master}. After correcting for scattered-light,
the master fibre flat frame is extracted following the same procedure used for the science spectra, and then
each extracted order of the flat frame is normalised by its maximum. This normalised flat spectrum is assumed
to represent an approximation of the blaze function. The correction for the blaze function of the science spectra
is performed by dividing the extracted spectrum of each order by the corresponding normalised flat spectrum.
This procedure also corrects for the column-averaged pixel-to-pixel variations of the CCD.

For echelle spectrographs using slits, CERES estimates the blaze function by using the spectrum of a hot
and rapidly rotating star where a high order polynomial is fitted to the spectrum of each echelle order.

However, for both type of spectrographs, given the differences in the continuum emission of the observed star and the
blaze correcting object (flat lamp or rapidly rotating star), the deblazed spectra will end up having a particular slope or
smooth variation that will not reflect the shape of the continuum of the observed star. The recovery of the stellar continuum
can be only found if the spectral energy distribution of the calibrator is accurately known, and flux calibration of echelle
spectrographs is notoriously difficult \citep[e.g.][]{suzuki:2003}.
However, as was already explained, the main goal of the CERES pipelines is to use of the narrow absorption lines, and therefore
the exact shape of the stellar continuum is not required. For this reason the sixth and seventh entries of the output of the CERES
pipelines correspond to the continuum normalized flux and its associated inverse variance, respectively (bottom panel of Figure~\ref{output}). 
The continuum normalisation is obtained by fitting a low order polynomial to the deblazed flux including an iterative procedure that
excludes absorption lines from the fit.
Finally the eighth column corresponds to the SNR per pixel in the continuum of the observed spectrum.  The SNR is estimated
by combining the contribution of (i) the expected Poisson error at each pixel given the extracted flux, (ii) the read out noise, and (iii) the expected Poisson error
at each pixel produced by the scattered light.

In addition to the reduced and wavelength calibrated spectra CERES contains several functions that are used
by the pipelines to analyse the spectra in an homogeneous way, as we now detail.

\subsubsection{Radial Velocities}
\label{sec:rvs}
The information about the velocity of the observed star is contained in the wavelength position of its spectral lines.
\cite{griffin:1967} showed that a very efficient way for using simultaneously the information provided by all the observed lines 
for measuring precise radial velocities is to use the cross-correlation function (CCF), which can be used even with
low signal-to-noise ratio data.
Following this principle, CERES includes a set of functions for the computation of the CCF using a binary mask 
\citep{baranne:1996}. The mask takes values equal to 1 in the regions where a typical stellar spectra
contains narrow absorption lines and equal to 0 elsewhere. The exact regions of the mask associated with absorption lines
will depend on the atmospheric properties of the star (principally on $T_{\rm eff}$). For this reason there are 3 available
masks for 3 different spectral types (G2, K5 and M5) that can be used to compute the RV of the observed spectrum.
These masks are the same ones used by the data reduction system of HARPS \citep{mayor:2003}.
The default mask for all the pipelines is the G2 but it can be changed by the user if the properties
of the observed star are known and are closer to the ones of the other masks. However, we note that it is 
key to analyse all the spectra of a particular star with the same mask, because the use of different
masks will produce different zero point velocities that can be responsible of increasing the dispersion of the
radial velocity measurements, hindering the detection of a Keplerian variation.

\begin{figure*}
\plotone{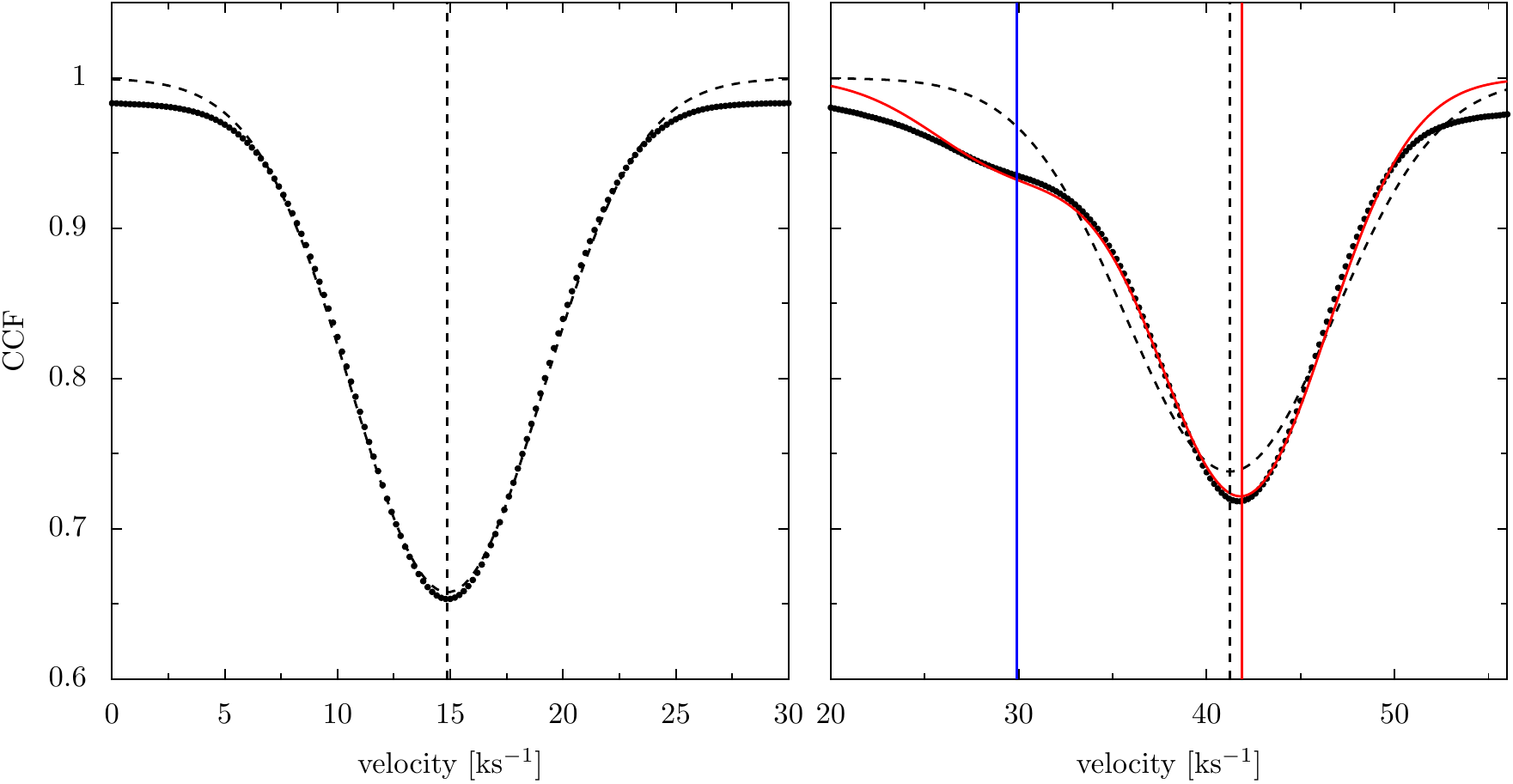}
\caption{Left panel: peak of the CCF obtained from a FEROS spectrum of a solar type star and the G2 binary mask. The dotted line correspond to the CCF and the dashed line to the Gaussian fit. The vertical dashed line correspond to the mean of the Gaussian and the corresponding radial velocity value computed for this spectrum. Right panel: peak of the CCF obtained for a fainter star in the presence of significant scattered moonlight. The velocity position of the secondary peak in the CCF produced by the moon is marked by the vertical blue line. In this case the simple Gaussian fit does not deliver a reliable estimation of the radial velocity and a double gaussian fit is applied (red lines).
 \label{ccfs}}
\end{figure*}

The algorithm that computes the CCF includes the effects of pixelization and computes one CCF for each echelle
order before combining them using a weighted sum according to the median SNR of each order. The CCF for the
echelle order $m$ at a certain velocity $v$ can be expressed as
\begin{equation}
CCF^m(v) = \frac{\int_{\lambda_i}^{\lambda_f} W(\lambda') F(\lambda) M(\lambda') d\lambda} {\sqrt{\int_{\lambda_i}^{\lambda_f}W(\lambda')M(\lambda')^2}},
\label{ccf}
\end{equation}
where $\lambda_i$ and $\lambda_f$ are the wavelengths of the initial and final pixels of the order $m$, $F(\lambda)$ corresponds
to the observed spectrum, $W(\lambda')$ corresponds to the weight that each spectral zone has according to the binary mask
and $M(\lambda')$ is the binary mask shifted to the $\lambda' = \lambda (1 + v/c)$ wavelength positions
due to the Doppler displacement, where $c$ is the speed of light.
As defined in equation \ref{ccf}, the CCF will acquire its minimum values for $v$ close to the radial velocity of the
observed star. The actual RV is computed by fitting a Gaussian to the CCF, and the resulting mean is taken to be the
RV of the star. The left panel of Figure~\ref{ccfs} shows an example of a CCF computed from a FEROS spectrum and its
corresponding fitted Gaussian.

During nights with a bright Moon, the spectrum of the Sun reflected by the Moon can strongly contaminate
the spectrum of the target star producing an additional peak in the CCF that can systematically
alter the measured radial velocity of the star. Different methods for correcting for this effect have been
proposed in the recent years \citep[e.g.][]{cabrera:2010,halverson:2016}.
CERES also includes tools for determining the impact that moonlight
contamination produces on the radial velocities. Using the \texttt{pyephem} package we compute the coordinates that the
moon possesses at the time of the observations and we also estimate the radial velocity at which the solar spectrum
should show up in the CCF. In addition, from previous simulations we also know the width of the CCF peak
produced by this contaminant. If the Moon is above the horizon and produces an important peak in the
CCF close to the one produced by the star, then we simultaneously fit two Gaussians to the computed CCF,
where the Gaussian corresponding to reflected sunlight has only one free parameter, namely the depth of the peak produced by the moonlight
contamination. The right panel of Figure~\ref{ccfs} shows a CCF that is heavily contaminated and the corresponding best fit
of Gaussians.

Uncertainties on the radial velocities are determined from the width of the CCF and the mean signal-to-noise ratio close to the
Mg triplet zone using empirical scaling relations \citep[as in][]{queloz:1995} whose parameters are
determined using Monte Carlo simulations. The exact equation used to estimate the RV error is
\begin{equation}
\label{rvsim}
\sigma_{RV} = b + \frac{a(1.6+0.2\sigma_{ccf})}{SN_{5130}},
\end{equation}
where $a$ and $b$ are the coefficients obtained via the Monte Carlo simulations and depend on the applied mask,
$SN_{5130}$ is the continuum SNR at 5130 \AA\ and $\sigma_{ccf}$ is the dispersion of the Gaussian fit
to the CCF.

Our Monte Carlo simulations were performed for every spectrograph that is supported by CERES.
We use high SNR extracted spectra, to which artificial Gaussian noise is added. The amplitude
of the noise varies along each echelle order by taking the blaze function into account. This artificial
spectrum is then continuum normalised and we compute its radial velocity using equation \ref{ccf}.
For a given noise amplitude, this process is repeated $\sim1000$ times and we save the standard deviation ($\sigma_{RV}$)
of the sample of computed radial velocities. This procedure is repeated with different noise amplitudes in order
to have an estimation of $\sigma_{RV}$ for different values SNR. In addition we use spectra with different
stellar atmospheric parameters which produce different widths of the cross-correlation peak ($\sigma_{ccf}$).
The sets of SNR, $\sigma_{RV}$ and $\sigma_{ccf}$ obtained with these simulation were then fitted with
the function specified in equation \ref{rvsim} in order to determine the appropriate $a$ and $b$ coefficients for each instrument.
For illustration, the values of the coefficients for the G2 mask in the case of the Coralie spectrograph 
are $a = 0.06544$ and $b = 0.00146$ (in km s$^{-1}$).

In order to measure the RV variations in a star produced by a planetary companion it is required
to correct the RV of the observed spectra for the velocity that the observer has with respect to the star.
This procedure is known as the barycentric correction and consists in computing the velocity that the
observatory has with respect to the barycentre of the solar system projected in the direction of the
observed star. The two principal velocities that have to be computed are the movement of the earth
around the barycentre and the rotation of the earth at the geographical coordinates of the observatory.
In order to determine the barycentric correction CERES uses the
\texttt{jplephem}\footnote{http://www.cv.nrao.edu/$\sim$rfisher/Python/py\_solar\_system.html}
python client of the Jet Propulsion Laboratory ephemerides package which is written
mostly in \texttt{C} and \texttt{Fortran} but can be called directly from \texttt{Python}. 
This velocity correction is computed by taking into account the integration time of the observation.
The default procedure is to use the mid-time of the exposure to compute the barycentric
velocity correction. However, if the spectrograph is equipped with an exposure meter, the time at which
the detector has received half of the total flux is used as the input of the \texttt{jplephem} package.

\subsubsection{Bisector Spans}
\label{ssec:BS}
\begin{figure*}
\plotone{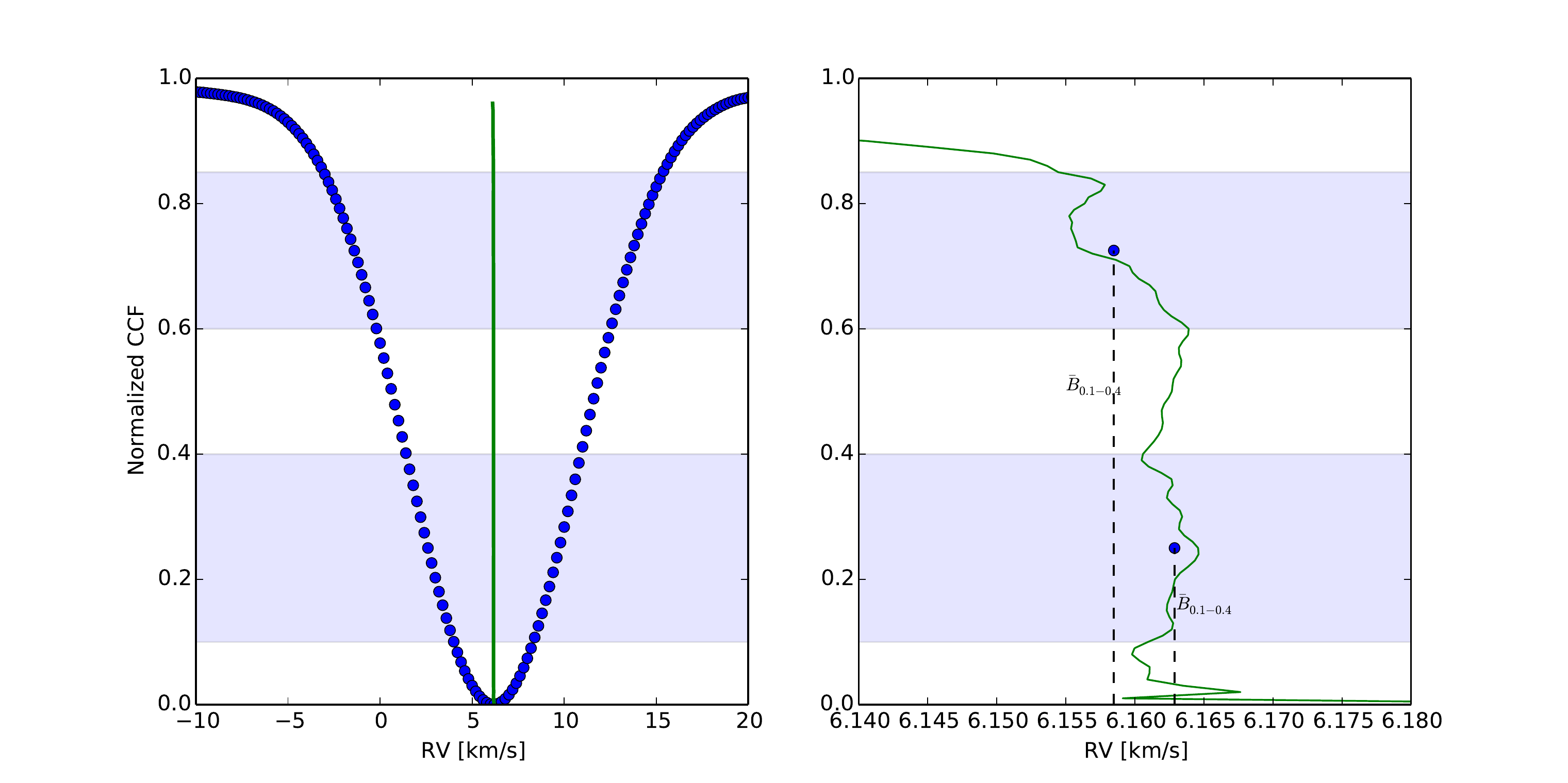}
\caption{Left panel: the blue points correspond to the computed CCF, while the green line traces the bisector values at different depths. The coloured
regions correspond to the two zones where the median bisectors are computed. Right panel: Zoom-in of the bisector values shown in the left panel, where the asymmetry of the CCF peak can be identified.
 \label{bss}}
\end{figure*}

Another useful parameter that can be extracted from the CCF is the bisector span (BS) which is a measure
of the asymmetry of absorption lines. The absorption lines can be naturally asymmetric due to the Doppler
broadening produced by convective motions in the surface of the star (macro turbulence). This effect
produces the granulation of the stellar surface where the luminosity of the raising material
is greater than the one of the material that is entering back the interior of the star. The observed spectrum
corresponds to the disk integrated stellar intensity and the effect of the different intensities in the rising and receding
material leads to the formation of asymmetric absorption lines. However, if the star has low levels of activity, 
the degree of asymmetry produced by convection
is constant in time. Variations in the measured bisector spans can lead to variations in the measured
radial velocities that can be associated with false positive scenarios. For example, eclipsing binary systems in which
both peaks cannot be resolved in the CCF will produce this kind of behaviour in the single CCF peak.
Similarly, stellar activity (spots and plages) leads to time correlated BS variations \citep[see e.g.][]{desort:2007, boisse:2011,dumusque:2014}.
In order to confirm the planetary nature of a TEP candidate, radial velocity and BS variations must be uncorrelated.
CERES computes the BS from the CCF following \cite{queloz:2001}. The bisector is computed as the mean
velocity $B(d)$ at the depth $d$ between both sides of the CCF peak as
\begin{equation}
B(d) = \frac{v_l(d) + v_r(d)}{2},
\end{equation} 
where, as shown in Figure~\ref{bss}, $v_l$ correspond to the velocities located on the left side from the minimum of the CCF peak
and $v_r$ are the ones on the right side. Then, the mean bisector is computed at two depths ranges:
from $d=0.1$ to $d=0.4$ and from $d=0.6$ to $d=0.85$ obtaining
\begin{equation}
\bar{B}_{0.1-0.4} = E[B(d)] , \forall\ 0.1 < d < 0.4,
\end{equation}
\begin{equation}
\bar{B}_{0.6-0.85} = E[B(d)] , \forall\ 0.6 < d < 0.85.
\end{equation}
Finally, the BS value is obtained from the difference of both quantities as 
\begin{equation}
BS = \bar{B}_{0.1-0.4} - \bar{B}_{0.6-0.85}.
\end{equation}
Figure~\ref{bss} shows the parameters that are extracted from the CCF for computing the bisector span.

The errors in the BS measurements are estimated using a scaling relation that depends on
the SNR in the region of the Mg triplet only as
\begin{equation}
\sigma_{BS} = \frac{a}{SN_{5130}} + b,
\end{equation}
where $a$ and $b$ are coefficients determined from Monte Carlo simulations, with random noise
added to high signal-to-noise ratio spectra for different spectral types, as was explained in \S~\ref{sec:rvs}.

\subsubsection{Rough stellar classification}

CERES also includes a tool for fast estimation of stellar atmospheric
parameters of the observed spectra.
In order to estimate the stellar parameters ($T_{\rm eff}$, $\log(g)$, [Fe/H], $v \sin i$) we cross-correlate
the observed spectrum against a grid of synthetic spectra of late-type stars from \cite{coelho:2005}
convolved to the resolution of the spectrograph and a set of $v \sin i$ values given by
{0, 2.5, 5, 7.5, 10, 15,..., 45, 50} km s${-1}$. First, a set of atmospheric
parameters is determined as a starting point by ignoring
any rotation and searching for the model that produces the highest
cross-correlation using only the Mg triplet region. We use
a coarse grid of models for this initial step ($\Delta T_{\rm eff} = 1000\ K$,
$\Delta \log(g) = 1.0$, $\Delta$[Fe/H] = 1.0). A new CCF between the observed spectrum and the model with
the starting point parameters is then computed using a wider
spectral region ($4800-6200 \ \AA$). From this CCF we estimate radial velocity
and $v\sin i$ values using the peak position and width of the CCF,
respectively. New $T_{\rm eff}$, $\log(g)$, and [Fe/H] values are then estimated
by fixing $v \sin i$ to the closest value present in our grid and
searching for the set of parameters that gives the highest cross-correlation,
and the new CCF is used to estimate new values of
RV and $v \sin i$. This procedure is continued until convergence on
stellar parameters is reached. For high signal-to-noise ratio spectra, uncertainties
of this procedure are typically $200\ K$ in $T_{eff}$, $0.3$ dex in $log(g)$,
$0.2$ dex in [Fe/H], and $2\ km s^{-1}$ in $v \sin i$. These uncertainties
were estimated from observations of stars with known stellar
parameters from the literature. Results of our stellar parameter
estimation procedure are obtained in $\approx$2 minutes using a
standard laptop.

This tool is specially useful in the context of reconnaissance spectroscopy
during the follow-up observations of transiting candidates, because
it allows to make efficient use of the available observing time by 
quickly identifying some false positives or troublesome systems,
such as fast rotators and giants, which are immediately dropped
out from the target list. However, this module not expected to
deliver high precision stellar parameters at publication level. For that purpose
we have developed an alternative code \citep[ZASPE][]{zaspe} that computes
precise stellar atmospheric parameters with reliable errors, and uses as input the
processed spectra obtained with CERES.

\label{sec:post:ceres}

\section{Currently Supported Instruments}
\label{instruments}

\begin{deluxetable*}{lccccccr}[!ht]
 \tablecaption{Properties of the instruments currently supported by CERES.}
\tablehead{
     Name &  Telescope &  Type & Resolution & Sampling & Thermal & Vacuum & CERES RV precision\\
     \colhead{} & {} & {}  & {$\times 1000$} & {px} & stability & vessel &  \colhead{m s$^{-1}$}
}
\startdata
\label{tab:rvs}
 ARCES         & APO 3.5m        & Slit    &$ 30           $&$ 2.8                 $ & No  & No & Unknown\\
 CAFE            & CAHA 2.2m     & Fibre &$ 70           $&$ 2.2                 $ & Yes & No & 20 \\
 Coralie          & Euler 1.2m      & Fibre &$ 60            $&$ 3.5                 $ & Yes & Yes & 7.8\\
 FEROS         & MPG 2.2m      & Fibre &$ 50            $&$ 2.2                 $ & Yes (200 mK) & No & 7.5\\
 FIES	     & NOT 2.5m        & Fibre &$ 25, 46, 67 $&$  6.4, 3.4, 2.4 $ & Tes (200 mK) & No & Unknown\\
 FIDEOS        & ESO 1.0m       & Fibre &$ 40            $&$ 2.7                 $ & Yes & No & Unknown\\
 HARPS         & ESO 3.6m       & Fibre &$ 85, 120    $&$ 4.3, 3.0          $ & Yes (10 mK) & Yes & 2.5 \\
 HIRES          & Keck 10m        & Slit    &$ 36, 48, 72 $&$ 6.3, 4.7, 3.1   $ & Yes & No & Unknown\\
 ECHELLE     & Du Pont 2.5m  & Slit    &$ 30 - 50     $&$ 2.3                 $ & No  & No & 500\\
 ESPaDOnS  & CFHT 3.6m      & Fibre &$ 80            $&$ 3.5 - 2.1         $ & Yes (100 mK)& Yes & Unknown\\
 MIKE            & Magellan 6.5m & Slit    &$ 20 - 80     $&$ 7.1 - 1.8         $ & No & No & Unknown\\
 PFS              & Magellan 6.5m & Slit    &$ 38 - 190   $&$ 4.9 - 1.0         $ & Yes & No & Unknown\\
 PUCHEROS & PUC 0.5m       & Fibre &$ 20            $&$ 2.8                  $ & No & No & 200\\
\enddata
 \label{table:inst}
\end{deluxetable*}

Currently, CERES has been used to build fully automated reduction and analysis pipelines for thirteen echelle spectrographs, ranging 
from low-mid ($R\approx 20000$) to high ($R\approx 120000$) resolution spectrographs.
A brief description of these instruments is given in the following paragraphs, while some specification can be found in Table \ref{table:inst}.
\begin{itemize}
\item{CORALIE}: This is a R=60,000  echelle spectrograph installed at the 1.2m Euler/Swiss telescope at La Silla Observatory.
It is fibre-fed from the telescope to a 2Kx2K CCD and includes also a second fibre that can be used to acquire a simultaneous
spectrum of the sky or of a wavelength comparison source (ThAr lamp or Fabry-Perot etalon). It is placed in a fully controlled room
where the temperature and vibrational conditions are stabilised.
The current CERES pipeline is able to handle the ThAr and Fabry-Perot simultaneous calibration modes.
The CERES pipeline developed for this instrument was briefly described in \citet{jordan:2014}
\item{FEROS} \citep{kaufer:1998}: The Fibre-fed Extended Range Optical Spectrograph is installed at the 2.2m MPG telescope in La Silla Observatory
and delivers a spectral resolution of $R\approx 50000$ by using an image slicer. It has 39 orders which are registered by a 4Kx2K CCD
and contains also a comparison fibre that can be used to obtain a simultaneous spectrum of the sky or a simultaneous ThAr lamp.
It is placed in a separate room where the environmental conditions are monitored and stabilised.
%Our pipeline is able to process data with the simultaneous ThAr lamp achieving an RV precision of  $\sigma_{RV}\approx 5$ m s$^{-1}$ on bright targets as opposed to the $\sigma_{RV} > 30$ m s$^{-1}$ obtained with the official pipeline.
\item{HARPS} \citep{queloz:2001:harps}: The  High Accuracy Radial velocity Planet Searcher is an echelle spectrograph installed at the 3.6m telescope
in La Silla Observatory. It is one of the most ambient-stabilised fibre fed spectrographs in the world and is capable of detecting planets
a few times more massive than the Earth. Its spectral resolution is $R=120000$ and it possesses 72 echelle orders which are
registered by a 2 chip mosaic detector of total size 4Kx4K. It is housed in a vacuum vessel to avoid spectral drifts due to temperature and air pressure variations which produce instrumental variations $<1$ m s$^{-1}$ along the night.
It also possesses a second fibre that can be used to monitor the background sky or to obtain a simultaneous wavelength calibration.
\item{CAFE} \citep{aceituno:2013}: The Calar Alto Fiber-fed Echelle spectrograph is installed in the 2.2m telescope at the Calar Alto Observatory in Spain.
It is placed in a temperature and vibration controlled room but it does not count with a simultaneous calibration
system. The resolution of this instrument is $R\approx 70000$ and the complete optical spectrum is divided in 84 orders and registered
in a 2Kx2K CCD.
%Due to the absence of a comparison fibre, 
The radial velocity precision of the CERES pipeline for this instrument is
$\sigma_{RV}\approx 30$ m s$^{-1}$.
\item{ESPaDOnS} \citep{donati:2003}: The Echelle SpectroPolarimetric Device for the Observation of Stars is installed at the 3.6m Canada-France-Hawaii Telescope in Manua Kea. It is a 
fiber-fed spectrograph that uses an image slicer to achieve a resolution of of 70000 and allows also to measure simultaneously circular and linear polarisations. The CERES pipeline can
process ESPaDOnS data obtained with the three different observing modes (star, star + sky, polarimetry).
\item{PUCHEROS} \citep{vanzi:2012}: This instrument is a $R=20000$ fibre fed echelle spectrograph built by the Centro de Astroingenieria at PUC/Chile and installed at the 0.5m telescope of the Observatorio UC near Santiago. It is not placed in a controlled room and
does not count with a simultaneous wavelength calibration system.
The CERES pipeline is the reduction system for this instrument and is installed at the observatory.
The small aperture of the telescope translates in that most of the observations have low SNR for which the optimal extraction
of the CERES pipeline plays a fundamental role in delivering good quality data.
\item{FIES} \citep{fies}:  the Fiber-fed Echelle Spectrograph is installed at the 2.5m Nordic Optical Telescope. It has three different fibres which deliver resolutions of 25000,
46000 and 67000, respectively. It also has a fourth fibre that can be used to obtain to perform a simultaneous wavelength calibration. The current version of the pipeline can
process only data obtained with the low-res and high-res fibres without simultaneous calibration.
\item{FIDEOS} \citep{tala:2014}:The FIbre Dual Echelle Optical Spectrograph was recently installed on the 1m telescope at La Silla Observatory.
Its object fibre has a image slicer which deliver a resolution of $R=40000$. It has also a comparison fibre that can be used to perform simultaneous wavelength calibration. Given that the instrument is still in the commissioning stage, the current CERES pipeline will probably be updated.
\item{DuPont/Echelle}: This instrument is an echelle spectrograph mounted on the 2.5m DuPont telescope in Las Campanas Observatory.
%Currently, the CERES pipeline is only able to process data acquired with the 1" slit which delivers a resolution of R=40,000.
This spectrograph is very unstable and the flexures of the telescope produce that the vertical and horizontal positions of the orders
change with the position of the telescope. However the pipeline is able to retrace the orders and deal with these stability related issues.
%The RV precision of the pipeline is $\sigma_{RV}\approx 200$ m s$^{-1}$.
\item{MIKE} \citep{bernstein:2003}: The Magellan Inamori Kyocera Echelle spectrograph is mounted on the 6.5m Clay telescope at Las Campanas Observatory. The spectra is separated into two arms and both portions are registered in different CCDs. 
Currently the CERES pipeline can only process the red arm (4900\ \AA-9500\ \AA) with
the $1\arcsec$ slit that delivers a resolution of $R=65000$. This instrument is also mounted directly on the telescope which strongly affects the achievable RV precision.
%The actual RV precision is unknown due to the lack of enough observations of RV standard stars.
\item{PFS} \citep{crane:2006}: The Carnegie Planet Finder Spectrograph is also mounted on the Clay telescope but this instrument uses the I$_2$ cell technique for achieving precise radial velocity measurements. The official pipeline developed and managed by the PFS team delivers
a radial velocity precision of $\sigma_{RV}\approx 2$ m s$^{-1}$ for bright stars. The CERES pipeline does not handle data with I$_2$ cell
and for the moment it only relies on the wavelength calibration provided by a ThAr lamp.
\item{ARCES} \citep{wang:2003}: The ARC Echelle Spectrograph is mounted on the 3.5m telescope at the Apache Point Observatory in New Mexico, USA, and has a spectral resolution of $R=31000$. 
\item{HIRES} \citep{vogt:1994}: The High Resolution Echelle Spectrometer is mounted on Keck in Hawaii.
The complete optical spectrum is registered in 3 CCDs but currently, the CERES pipeline can only reduce data of the green chip (5000\ \AA-6000\ \AA)
and does not compute radial velocities with the I$_2$-cell technique.
\end{itemize}

\section{Performance and interesting results}
\label{results}

The computational routines and recipes developed for CERES were designed with the main goal of performing spectroscopic follow-up
observations of transiting planetary candidates from the HATSouth survey \citep{bakos:2009}.
Therefore, the tools we have presented were optimised for obtaining homogeneous results from low to moderate signal to noise ratio data (SNR $\sim$ 20-50),
including radial velocities precise enough to measure variations produced by massive planetary companions (5-100 m s$^{-1}$) and atmospheric
parameters with enough precision to have an informed guess about the physical parameters of the observed stars.

\begin{figure*}
\plotone{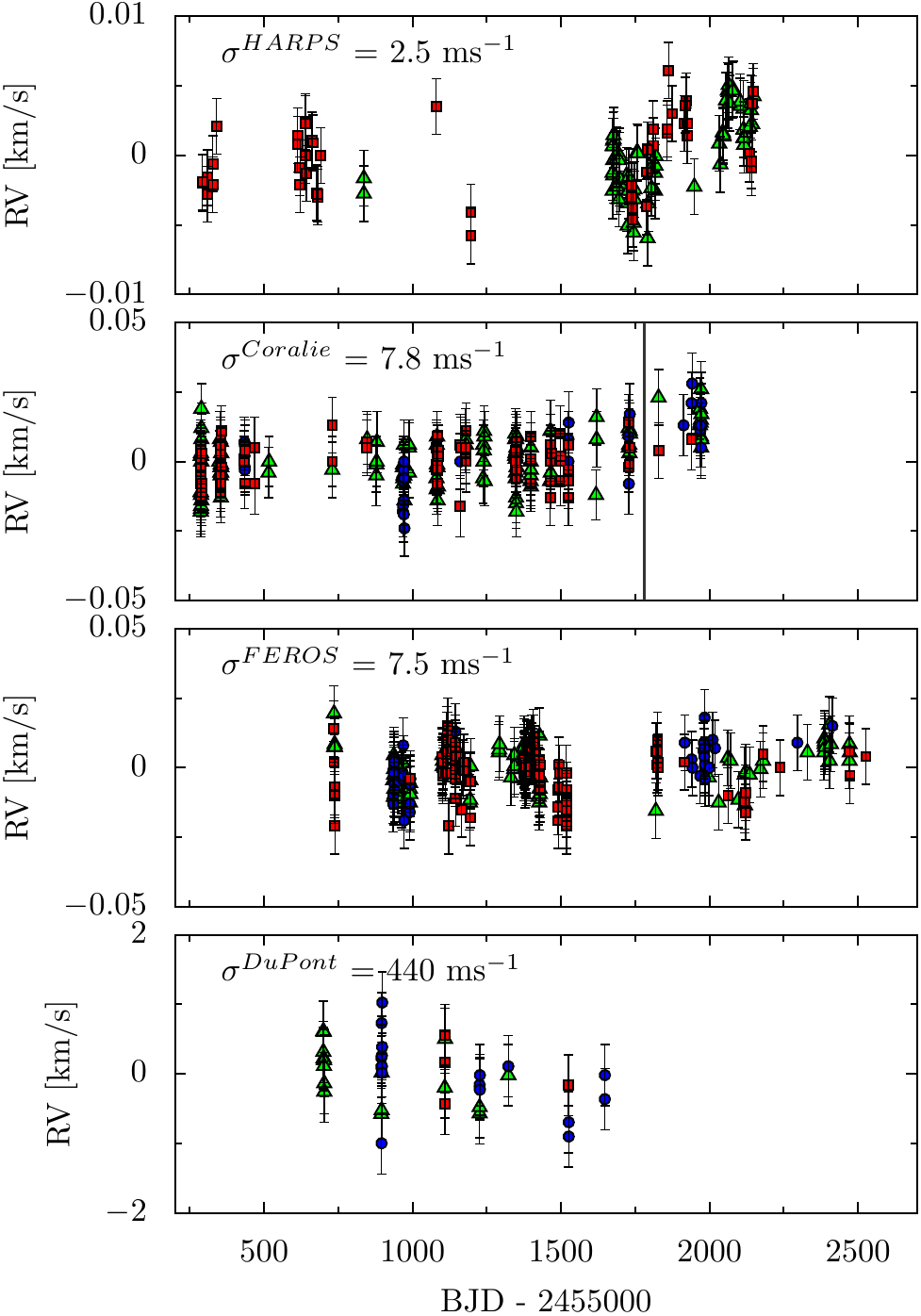}
\caption{Radial velocity measurements for 3 radial velocity standard stars (squares: HD72673, triangles: HD157347, circles: HD32147) obtained with four different instruments (from top to bottom: HARPS, CORALIE, FEROS, DuPont) using the CERES pipelines. The vertical line in the second panel corresponds to the date where the instrument was upgraded, which produced a $\sim$20 ms$^{-1}$ offset in the radial velocities.
 \label{rest}}
\end{figure*}

We have been monitoring the radial velocity precision of our pipelines for each spectrograph by observing radial velocity standard stars.
Figure~\ref{rest} shows the radial velocity measurements for these standard stars in the case of four different instruments,and Table \ref{tab:rvs}
 lists the individual dispersion values of the measured radial velocities for each spectrograph.
Figure~\ref{rest} shows that while the long term precision achieved by fibre-fed and stabilised instruments is below 10 m s$^{-1}$, the precision obtained
by the non stabilised spectrograph is of the order of 400 m s$^{-1}$, which is not enough for detecting variations produced by planetary mass
companions, but is sufficient for quickly identifying stellar or brown dwarf companions, which are common false positive systems in
dedicated ground-based surveys of transiting planets. Given that the CERES pipelines are also capable of estimating stellar atmospheric parameters
and computing bisector spans, the spectra obtained with the DuPont echelle can be used to efficiently identify other kind of false positives,
namely, giant stars, fast rotators and blended eclipsing binaries. The automatisation of the reduction steps make of the CERES pipeline
for the duPont echelle an optimal system for performing the reconnaissance spectroscopy steps of the follow-up observations of transiting
surveys, specially because there is no dedicated reduction system for this instrument. The CERES pipeline for the DuPont/echelle has been only tested for 2 slit configurations, namely $1\arcsec \times4\arcsec$ and 0.75$\arcsec\times 4\arcsec$, which
correspond to spectral resolutions of $R\approx 40000$ and $R \approx 50000$, respectively.

\begin{deluxetable}{lcccc}[!ht]
 \tablecaption{Properties of the radial velocity standard stars.}
\tablehead{
     ID &  Spectral type &  $\sigma_{RV}^{coralie}$ & $\sigma_{RV}^{feros}$ & $\sigma_{RV}^{harps}$ \\
     \colhead{} & {}  & \colhead{m s$^{-1}$} & \colhead{m s$^{-1}$} & \colhead{m s$^{-1}$}
}
\startdata
\label{tab:rvs}
 HD72673   & G9V &$ 8.0 $&$ 7.3 $&$ 2.6   $\\
 HD157347 & G3V &$ 7.3 $&$ 7.9 $&$ 2.3   $\\
 HD32147   & K3V &$ 8.5 $&$ 8.0 $&$ -  $\\
\enddata
 \label{table:rv_list}
\end{deluxetable}

The results obtained with the CERES pipelines for stabilised fibre-fed instruments (CORALIE, FEROS and HARPS), are precise enough for
allowing the detection of planetary mass companions. The accuracy obtained for the HARPS spectrograph is slightly worse than the one
obtained by the official Data Reduction Software, which is expected given the large amount of effort and numerous upgrades that have been
applied on the official dedicated pipeline.
The radial velocity precision achieved in the case of the CORALIE and FEROS pipelines is of $\approx$7 m/s. This precision is similar
to the one obtained by the official (non-public) CORALIE pipeline, but is significantly better than the $\approx30$ m s$^{-1}$ precision achieved by the dedicated FEROS pipeline installed at the telescope. In addition to the high precision achieved in the high signal to noise ratio regime, the FEROS pipeline based on CERES
routines can handle data with low signal to noise ratio (SNR $<$30) obtaining radial velocity uncertainties governed by Poisson errors, as opposed to the
official pipeline, whose velocity errors become dominated by systematics that are not removed in the reduction steps. A strong proof of
the capabilities of our CERES pipeline for FEROS is the discovery of HATS-20b \citep{bhatti:2016}, a Saturn-mass transiting planet around a 
$V=14$ solar type star, that produces a $K=40$ m s$^{-1}$ keplerian signal that was identified using 10 FEROS exposures of 1800s.
An important peculiarity of our CERES pipeline for FEROS is that it does not process
all the echelle orders because we have found that the precision of the wavelength solution significantly worsens
if all of them are included in the fit. Therefore, our final FEROS output has 25 echelle orders covering only from 3800 \AA\ to 6800 \AA.

These CERES pipelines developed for FEROS and CORALIE have been fundamental for confirming the planetary nature of most
of the discovered transiting planets from the HATSouth survey \citep[see e.g.][]{brahm:2015, mancini:2015, rabus:2016, ciceri:2016, brahm:2016}, and equally important to reject a much larger fraction of false positives.

Due to the high precision and homogeneity of the obtained results, the CERES pipelines, in particular the ones for CORALIE and FEROS have been already
used in different studies with different scientific goals, which we briefly mention here as a demontration of their scope:
\begin{itemize}
\item Kepler K2 planets: \citet{espinoza:2016} presented the discovery of a dense Neptune mass planet using data from the CORALIE and HARPS spectrographs. \citet{brahm:2016b} presented the discovery of two hot Jupiters, firstly identified as transiting candidates from K2 photometry, and which
planetary nature was confirmed via precision radial velocities from the CORALIE, FEROS and HARPS spectrographs. \citet{crossfield:2016} presented the validation of 104 planets from K2, for which FEROS observations reduced with CERES were used.
\item Radial velocity planets: Using HARPS and CORALIE data, \citet{jenkins:2016} announced the discovery of eight new planetary systems containing giant planets. \citet{wittenmyer:2016} and \citet{jones:2016} presented the discovery of five new systems conformed of giant planets orbiting giant stars by using data from the FEROS spectrograph among others.
\item Studies on already discovered extrasolar plates: \citet{ciceri:2016b} and \citet{mancini:2016} updated the physical parameters of the WASP-45b, WASP-46 and WASP-98 planetary systems by obtaining new spectroscopic and photometric observations of these transiting hot Jupiters. The spectroscopic data was obtained with FEROS.
\item Detached eclipsing binaries: \citet{helminiak:2014} and \citet{coronado:2015} presented the orbital and physical parameters of two new detached eclipsing systems that were identified from ASSAS photometry, and then followed-up with the FEROS, CORALIE, HARPS and PUCHEROS spectrographs.
\item Search of supernova progenitors: Combining data from RAVE, GALEX, HST and FEROS, \citet{parsons:2015} reported the discovery the first pre-supersoft X-ray binary.
\item Identification of members of young associations: \citet{elliott:2016} presented the discovery of 84 low mass stars that are linked to young associations. A fraction of those new members were confirmed using radial velocities from FEROS reduced with CERES.
\item Spectroscopic follow-up of Novae: \citet{izzo:2015} presented the first detection of lithium produced in the optical spectra of a Nova using early CERES spectra from FEROS and PUCHEROS.
\end{itemize}

\section{Conclusions}
\label{sec:conclusions}

We have presented CERES, a set of routines that allow the development of robust and fully automated pipelines for the
reduction, processing and analysis of echelle spectra. We have constructed pipelines for thirteen different instruments with quite different specifications, reaching results that are almost as good or better than those of dedicated pipelines when available. In this regard, the CERES pipeline for the FEROS spectrograph stands out due to the high RV precision that can be achieved in the high signal-to-noise ration regimen ($\sigma_{RV}$=7.5 ms$^{-1}$), and also for its good behaviour for low signal-to-noise ratio data, which has allowed the discovery of several extrasolar planets orbiting stars even fainter than $V=14$. Moreover, we have developed reduction pipelines for instruments that do not have any dedicated pipeline, like the one for the DuPont spectrograph, for which we can achieve an RV precision of $\sigma_{RV}$=400 ms$^{-1}$ and complete automatization despite the lack of stability of the instrument. The recipes for these thirteen instruments can be used as guide for constructing pipelines for other instruments. In addition to reduced spectra, radial velocities and bisector spans, the CERES pipelines estimate rough and fast atmospheric parameters which are useful for quick target vetting at the telescope.

There are several limitations of CERES that are being considered for future upgrades. For example, our extraction algorithms do not perform a previous rectification of the curvature of the echelle orders. The correction for the curvature can increase the resolution of the extracted spectrum, but it could introduce complications for the optimal extraction algorithm because it will require an interpolation between pixels.
In addition, given that the principal applications of CERES have been focused on fibre-fed spectrograph, it does not contain a routine to correct for sky contamination, which can be useful in the case of slit spectrographs. Finally, an important upgrade for CERES will consist in developing automated routines for the computation of precise radial velocities using the iodine cell technique.

The full CERES code, along with the pipelines for the instruments described in Section \ref{instruments} have been made publicly available\footnote{https://github.com/rabrahm/ceres}.
\label{concl}

\acknowledgments

We thanks Gaspar Bakos and Joel Hartman for many useful discussions and strong encouragement to develop and improve the code presented in this work.
A.J. thanks Didier Queloz with whom he had many fruitful discussions regarding radial velocity measurements in the early development stages of CERES.
R.B. and N.E. are supported by CONICYT-PCHA/Doctorado Nacional. A.J.\ acknowledges support from FONDECYT project 1130857 and from BASAL CATA PFB-06. R.B., N.E. and  A.J. acknowledge support from the Ministry for the Economy, Development, and Tourism Programa Iniciativa Cient\'ifica Milenio through  grant IC 120009, awarded to the Millennium Institute of Astrophysics (MAS).
This paper made use of the SIMBAD database (operated at CDS, Strasbourg, France), NASA's Astrophysics Data System Bibliographic Services, and data products from the Two Micron All Sky Survey (2MASS) and the APASS database and the Digitized Sky Survey.

%\clearpage
\bibliographystyle{apj}
\bibliography{ceres}

\end{document}